\documentclass[conference]{IEEEtran}
\IEEEoverridecommandlockouts
\usepackage{cite}
\usepackage{amsmath,amssymb,amsfonts}
\usepackage{algorithmic}
\usepackage{graphicx}
\usepackage{textcomp}
\usepackage{multirow}
\usepackage[inline]{enumitem}

\usepackage{booktabs}
\usepackage{multirow}
\usepackage[table,xcdraw]{xcolor}

\usepackage[hyphens]{url}
\usepackage[hidelinks]{hyperref}
\usepackage[numbers]{natbib}

\def\BibTeX{{\rm B\kern-.05em{\sc i\kern-.025em b}\kern-.08em
    T\kern-.1667em\lower.7ex\hbox{E}\kern-.125emX}}

\begin{document}



\title{A Blockchain-based Data Governance Framework with Privacy Protection and Provenance for e-Prescription}






\author{
\IEEEauthorblockN{Rodrigo Dutra Garcia\IEEEauthorrefmark{1}, Gowri Sankar Ramachandran\IEEEauthorrefmark{2}, Raja Jurdak\IEEEauthorrefmark{2}, and Jo Ueyama\IEEEauthorrefmark{1}}
\IEEEauthorblockA{\IEEEauthorrefmark{1}\textit{Institute of Mathematics and Computer Science, University of São Paulo, Brazil.} \emph{\{rgarcia,joueyama@icmc.\}@usp.br}}
\IEEEauthorblockA{\IEEEauthorrefmark{2}\textit{School of Computer Science, Queensland University of Technology, Australia.}  \emph{\{g.ramachandran,r.jurdak\}@qut.edu.au}}
}


\maketitle

\begin{abstract}
Real-world applications in healthcare and supply chain domains produce, exchange, and share data in a multi-stakeholder environment. Data owners want to control their data and privacy in such settings. On the other hand, data consumers demand methods to understand when, how, and who produced the data. These requirements necessitate data governance frameworks that guarantee data provenance, privacy protection, and consent management. We introduce a decentralized data governance framework based on blockchain technology and proxy re-encryption to let data owners control and track their data through privacy-enhancing and consent management mechanisms. Besides, our framework allows the data consumers to understand data lineage through a blockchain-based provenance mechanism. We have used Digital e-prescription as the use case since it has multiple stakeholders and sensitive data while enabling the medical fraternity to manage patients' prescription data, involving patients as data owners, doctors and pharmacists as data consumers. Our proof-of-concept implementation and evaluation results based on CosmWasm, Ethereum, and pyUmbral PRE show that the proposed decentralized system guarantees transparency, privacy, and trust with minimal overhead.

\end{abstract}


\section{Introduction}

Prescription systems allow healthcare professionals, such as physicians, to create digital records about a patient's health status by adding diagnosis and medications data. It allows for more efficient communication and reduced inconsistencies compared to paper-based prescriptions \cite{aldughayfiq2021digital,samadbeik2017copmarative}. Thus, digital prescription systems increase globally, enabling multiple stakeholders, including doctors and pharmacies, to access and manage patients' data effectively. 

Patients want to control their data and privacy in healthcare settings since prescription and diagnosis data contain sensitive and personally identifiable information. Note that unauthorized parties may gain access and misuse patients' data \cite{id9-n5-patients-controlled}. Therefore, it is essential to \emph{protect patients' privacy while letting them manage and permit access to their data transparently}, which is one of the problems this paper aims to investigate. 

\begin{table*}[]
\centering
\bgroup
\setlength{\tabcolsep}{2em}
\def\arraystretch{1.6}
\caption{related work and gaps}
\label{tab:related-work}
\begin{tabular}{>{\kern-\tabcolsep}cccccccc<{\kern-\tabcolsep}}
\hline
 &
  \multicolumn{6}{c}{\textbf{Related Work}} &
  \multicolumn{1}{l}{} \\ \cline{2-7}
\multirow{-2}{*}{\textbf{Features}} &
  \cite{comp-id2-n1-secure-rx, comp-id2-n2-rxblock,  rodrigo-gabriel-gowri-jo} &
  \cite{id9-n2-preserving-patient-privacy} &
  \cite{id10-n2-SPChain} &
  \cite{id11-n4-a-new-blockchain} &
  \cite{id11-n9-DMMS} &
  \cite{MedBlock-KaiFan-2018, Towards-secure-2018} &
  \multicolumn{1}{l}{\multirow{-2}{*}{\textbf{This work}}} \\ \cline{1-1} \cline{8-8} 
Decentralization &
  \cellcolor[HTML]{BCEBBC}Y &
  \cellcolor[HTML]{BCEBBC}Y &
  \cellcolor[HTML]{BCEBBC}Y &
  \cellcolor[HTML]{BCEBBC}Y &
  \cellcolor[HTML]{BCEBBC}Y &
  \cellcolor[HTML]{BCEBBC}Y &
  \cellcolor[HTML]{BCEBBC}\textbf{Y} \\
e-Prescription &
  \cellcolor[HTML]{BCEBBC}Y &
  \cellcolor[HTML]{FFAFAF}N &
  \cellcolor[HTML]{FFAFAF}N &
  \cellcolor[HTML]{FFAFAF}N &
  \cellcolor[HTML]{BCEBBC}Y &
  \cellcolor[HTML]{FFAFAF}N &
  \cellcolor[HTML]{BCEBBC}\textbf{Y} \\
Privacy Support &
  \cellcolor[HTML]{FFAFAF}N &
  \cellcolor[HTML]{BCEBBC}Y &
  \cellcolor[HTML]{BCEBBC}Y &
  \cellcolor[HTML]{BCEBBC}Y &
  \cellcolor[HTML]{BCEBBC}Y &
  \cellcolor[HTML]{BCEBBC}Y &
  \cellcolor[HTML]{BCEBBC}\textbf{Y} \\
Patient Consent &
  \cellcolor[HTML]{FFAFAF}N &
  \cellcolor[HTML]{FFAFAF}N &
  \cellcolor[HTML]{FFAFAF}N &
  \cellcolor[HTML]{FFAFAF}N &
  \cellcolor[HTML]{FFAFAF}N &
  \cellcolor[HTML]{FFAFAF}N &
  \cellcolor[HTML]{BCEBBC}\textbf{Y} \\
Tracking Mechanism &
  \cellcolor[HTML]{FFAFAF}N &
  \cellcolor[HTML]{FFAFAF}N &
  \cellcolor[HTML]{FFAFAF}N &
  \cellcolor[HTML]{FFAFAF}N &
  \cellcolor[HTML]{FFAFAF}N &
  \cellcolor[HTML]{FFAFAF}N &
  \cellcolor[HTML]{BCEBBC}\textbf{Y} \\ \hline
\end{tabular}
}
\end{table*}

Pharmacies must sell certain drugs such as antibiotics with a valid doctor's prescription. A prescription containing an antibiotic medicine is valid for only a single purchase, meaning the pharmacy and the patient must obey the recommended dosages. However, pharmacies tend to sell medications illegally to patients even with the old and used prescription to gain financial revenue. Such illegal sales would lead to unwanted side effects, including drug abuse and overdoses~\cite{misuse-medication1,misuse-medication2}, burdening the healthcare system. Therefore, it is essential to regulate the medicine supply chain to prevent the unauthorized sales of medications, which is one of the focuses of this work.

Existing digital prescription systems primarily employ a centralized architecture, offering limited to no visibility into the operations providing maximum power to the administrating organization \cite{aldughayfiq2021digital,samadbeik2017copmarative}. Such centralized architectures are susceptible to single points of failure, enabling opportunities for data tampering. In addition, centralized systems may also misuse patients' health data without their consent, resulting in privacy violations. In summary, centralized architecture offer no transparency undermining the integrity of medical information while affecting patients' privacy \cite{security-health}. We, therefore, argue that a decentralized architecture with support for consent management, privacy preservation, data provenance, and compliance validation is essential for a trusted digital prescription system.

Existing works in digital e-prescription do not securely manage consent while providing support for privacy protection and accountability. Table~\ref{tab:related-work} summarizes the digital prescription literature and highlights the gap.




We propose a decentralized data governance framework for the electronic prescription that,

\begin{itemize}
\item Helps patients \emph{store, manage, and share} prescription data with other stakeholders through a tamper-proof ledger.
\item Protects patients' \emph{privacy} by storing encrypted prescription data on the blockchain ledger
\item Provides support for \emph{consent management} using proxy re-encryption scheme and smart contracts
\item Supports \emph{data provenance} to let data owners and data consumers efficiently monitor the historical records of the data and its origin, including who accessed the data and for what purposes
\item Uses \emph{privacy-enhancing data management} scheme to withhold personally identifiable and sensitive information from third parties, including drug regulators
\item Enables the drug regulators to control and monitor the flow of medications to the pharmacies through the \emph{accountable} blockchain ledger, thereby limiting illegal sales.
\end{itemize}
We have developed a proof-of-concept implementation using the CosmWasm, which uses Tendermint (a Byzantine Fault Tolerance (BFT) consensus mechanism), Ethereum ( a popular blockchain platform with smart contract support), and NuCypher pyUmbral \cite{nunez2018umbral} proxy re-encryption (PRE) library to estimate the overhead and feasibility. Our evaluation results show that the proposed data governance framework introduces minimal overhead while letting data owners control and manage their data with transparency and trust guarantees. Although we discuss the data governance framework through an e-prescription use case, the proposed framework is suitable for any multi-stakeholder application, including supply chain management, dealing with digital and sensitive data.

\section{Related Work}

Electronic prescription systems operate in a multi-stakeholder environments. It requires the integrity and transparency of information to avoid illegal drug sales while preventing patients' health problems due to drug overdose. Besides, the application of privacy-preserving techniques for medical records is another requirement to avoid misuse of sensitive information present in prescriptions. \citeauthor{comp-id2-n1-secure-rx} proposed SecureRx  \cite{comp-id2-n1-secure-rx}, a blockchain solution using the Ethereum platform to maintain patient records and prescriptions. Similarly, \citeauthor{comp-id2-n2-rxblock} suggested RxBlock \cite{comp-id2-n2-rxblock}, a solution using Ethereum to manage prescriptions and avoid drug overdose. \citeauthor{rodrigo-gabriel-gowri-jo} \cite{rodrigo-gabriel-gowri-jo} proposed a decentralized e-prescription system using smart-contracts on a BFT platform. However, these solutions do not \emph{manage consent} and focus on writing records to an immutable ledger without providing mechanisms to \emph{track who accessed the data and for what purposes} while \emph{protecting patient's sensitive information}.


Other research works investigate approaches to ensure the integrity and privacy of medical records by preventing tampering and data leakage. \citeauthor{id9-n2-preserving-patient-privacy} \cite{id9-n2-preserving-patient-privacy} presents a model to preserve patient data between different institutions using asymmetric encryption, particularly proxy re-encryption (PRE) on permissioned blockchain. Similarly, \citeauthor{id10-n2-SPChain} proposes SPchain \cite{id10-n2-SPChain}, a blockchain and PRE-based solution for sharing electronic health records (EHR). \citeauthor{id11-n4-a-new-blockchain} \cite{id11-n4-a-new-blockchain} investigate an electronic prescription system with the adoption of the k-anonymity method based on differential privacy for data protection. \citeauthor{id11-n9-DMMS} introduced DMMS \cite{id11-n9-DMMS}, a solution that exploits blockchain technology for medication history management and electronic prescriptions. \citeauthor{MedBlock-KaiFan-2018} proposes MedBlock \cite{MedBlock-KaiFan-2018} and \citeauthor{Towards-secure-2018} \cite{Towards-secure-2018} presents a blockchain-based solution for sharing medical records with privacy and threat analysis. Table \ref{tab:related-work} summarizes the related works and highlights the gap.

\section{Background on Proxy Re-Encryption (PRE)}

Proxy re-encryption is an asymmetric encryption technique initially proposed by Blaze et al. \cite{blaze} in which an entity $A$ (delegator) can delegate the decryption rights to another entity $B$ (delegatee) through a proxy server. In the PRE technique, we have three main actors:

\begin{itemize}
\item \textbf{Delegator}: data owner and delegates decryption rights to another user or entity (i.e., delegatee);

\item \textbf{Delegatee}: data consumer and receive decryption rights to access the encrypted information;

\item \textbf{Proxy}: performs a re-encryption using the delegation key (generated by the delegator) and the encrypted message to allow the delegatee to access the encrypted information;

\end{itemize}

Initially, a message $m$ is encrypted using the delegator's public key, $C_{A} = Enc(pk_{A}, m)$, and stored in a database. If delegatee $B$ needs to decrypt the message, he must initially request decryption rights for the delegator informing his public key $pk_{B}$. If the delegator agrees, it will produce a delegation key $rk_{A\to B}$ and send it to the proxy.

For delegatee $B$ to be able to decrypt the information, the proxy server must use the delegation key $rk_{A\to B}$ to re-encrypt $C_{A}$, that is, $C_{B} = ReEnc(rk_{A\to B}, C_{A})$. After re-encryption, the delegatee can use his private (i.e., secret) key $sk_{B}$ and decrypt the message. At all stages, only the public key is shared between the participants. From the proxy's point of view, it does not learn or try to decrypt confidential information. It receives encrypted information $C_{A}$ and sends other encrypted information $C_{B}$.




\section{Decentralized Architecture with Consent Management and Privacy Protection}

\subsection{System Model and Threats}

We assume a system comprising of patients, doctors, and pharmacies. When a patient visits a doctor, the doctor creates a new medical record that includes diagnostic data, personal details such as name and age, and prescriptions.

We focus on the following threats:
\begin{itemize}
\item{Privacy threat:} The patient's medical record includes sensitive data, which should not be revealed to unauthorized third parties without the patient's consent.
\item{Illegal drug sales:} The lack of visibility into the medication supply chain leads to illegal medication sales, resulting in drug overdoses.
\end{itemize}

Given these threats, this work aims to develop a solution with the following objectives.

\textbf{Objective 1:} We aim to develop a transparent medical prescription system based on the blockchain without revealing sensitive data to unauthorized third parties. Note that the data stored on the blockchain is visible to the public in a blockchain platform. \emph{Can we allow the patients to store and manage medical data in a tamper-proof ledger without violating patients' privacy?}

\textbf{Objective 2:} When the data get stored on a digital system, doctors and health care agencies can access the data for diagnostic and survey purposes. Under this circumstance, it is important to let patients or data owners have visibility into data usage. \emph{Can we allow the data owners to track and govern data usage by other parties?}

\textbf{Objective 3:} When a medication supply chain operates without a regulator, pharmacies can easily acquire drugs from the manufacturer and sell them to patients without a valid prescription. \emph{Can we prevent illegal drug sales by involving a regulator in a decentralized prescription system? To generalize it to other multi-stakeholder applications, can we allow the regulatory body to access data for compliance verification within a decentralized data governance infrastructure?}

\subsection{Proposed Solution}
We propose a decentralized data governance framework using blockchain technology and smart contracts to help patients manage their data more efficiently. When a patient consults a doctor, the doctor creates prescription data by recording the diagnosis, recommended medications, and dosage. Then, the prescription data is encrypted using the patient's public key and stored in the blockchain via the smart contract. We assume that the patient shares her public key with the doctor when she makes an appointment to see the doctor. The patient needs to allow the pharmacists access to the prescription data to receive medication from the pharmacy. We propose a data access tracking mechanism within the contract state to monitor data accesses. In this way, any query or update in the status of the records will be registered. Note that the existing blockchain-based systems support writing data to an immutable ledger. Still, they do not monitor or provide support for governing data usage, which is necessary for data provenance and privacy. Our framework not only records the data on an immutable ledger in a privacy-preserving manner but also logs access requests to govern data usage. 


To prevent illegal medications sales by the pharmacy, the regulatory agency will count, through a control contract, the number of drugs supplied to the pharmacy with the number of drugs sold (in the sales contract). Furthermore, we recommend a Crime Stoppers model for patients to report pharmacies' illegal actions, discussed in Section~\ref{sec:disc}. Patients can report the pharmacy that sells unlawful drugs and receive rewards (tokens) through a smart contract through this approach.

We assume that the blockchain can hold encrypted prescription data for brevity. We can extend the framework by storing the encrypted prescription data on off-chain storage while maintaining the hash on-chain, which we plan to tackle in our future work.



\subsection{Overview of Smart Contracts}
\label{sec:overview-smart-contracts}

Figure \ref{fig:stakeholders-and-contracts.pdf} shows the contracts among stakeholders for the proposed model:

\begin{figure}[h!]
\centering
\includegraphics[scale=0.50]{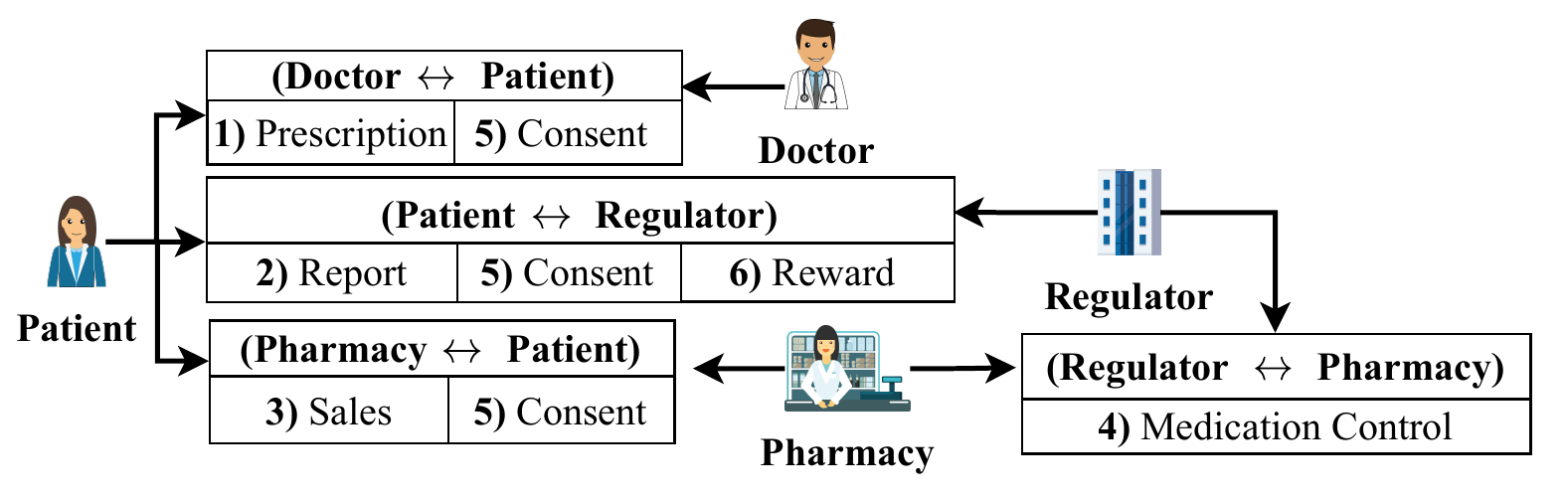}
\caption{Stakeholders and smart contracts}
\label{fig:stakeholders-and-contracts.pdf}
\end{figure}

\begin{enumerate}
    \item \textbf{Prescription Smart Contract}: 
the doctor creates the prescription data and invoke the \textit{create\_prescription} smart contract method. The contract state will be updated with patient's personal information, recommended medications, and diagnosis. Each instance has the address of the doctor (sender) and patient (recipient). The \textit{create\_prescription} method only accepts transactions from the doctor. In this sense, only transactions signed by the doctor will be valid. For data usage tracking, any query to the contract, the stakeholder address will be updated in the \textit{last\_access} state.

    \item \textbf{Report Smart Contract}: in case of illegal actions performed by the pharmacy, such as selling medication without a valid prescription, the patient can report it to the regulatory authority. If the report is valid, the patient can receive tokens as a reward.
The contract instance contains the source address and the destination (regulator). From the \textit{create\_report} method, the current contract state will be updated with the source address and the data (description) of the denunciation. Personally identifiable information (PII) will not be stored in the contract state.

    \item \textbf{Sales Smart Contract}: allows the pharmacy to sell medications to the patient. The pharmacy creates a contract instance with the patient to send sales transactions. Each sales transaction will be sent to the \textit{sell\_medication} contract method, and the current state is updated with the sales data (i.e., medication name, dosage, and price).

    \item \textbf{Medication Control Smart Contract}: allows the regulator to account for the number of drugs supplied and sold by the pharmacy. The legal amount will be sent by the regulator and received by the pharmacy. Only the pharmacy will notify the number of medications sold, and the smart contract method will automatically update the number of available medications. In this way, the regulator will account for the relationship between drugs supplied and drugs sold.

    \item \textbf{Patient Consent Smart Contract}: allows stakeholders to request the rights to decrypt the patient's prescription data. The contract state includes request origin address and the patient's consent, authorizing or not the data decryption. In Section     \ref{sec:request-smart-contract-delegation-key}, the request mechanism using smart contracts is presented.

    \item \textbf{Reward Smart Contract}: used for the regulator to transfer tokens to the patient in case of complaint proof. We plan to develop and detail the token economy in our future work.

\end{enumerate}

\begin{table}[]
\centering
\bgroup
\def\arraystretch{1.2}
\caption{Access to information by stakeholders after patient permission}
\label{tab:privacy-levels}
\begin{tabular}{>{\kern-\tabcolsep}cccc<{\kern-\tabcolsep}}
\hline
                   & \multicolumn{3}{c}{\textbf{Prescription Data}}                                          \\ \cline{2-4} 
\multirow{-2}{*}{\textbf{Stakeholder}} &
  \textbf{\begin{tabular}[c]{@{}c@{}}Personal Information\\ (e.g., Name, Age)\end{tabular}} &
  \textbf{Diagnosis} &
  \textbf{\begin{tabular}[c]{@{}c@{}}Medication\\  and  Dosage\end{tabular}} \\ \hline
\textbf{Doctor}    & \cellcolor[HTML]{BCEBBC}Yes & \cellcolor[HTML]{BCEBBC}Yes & \cellcolor[HTML]{BCEBBC}Yes \\
\textbf{Patient}   & \cellcolor[HTML]{BCEBBC}Yes & \cellcolor[HTML]{BCEBBC}Yes & \cellcolor[HTML]{BCEBBC}Yes \\
\textbf{Pharmacy}  & \cellcolor[HTML]{FFAFAF}No  & \cellcolor[HTML]{FFAFAF}No  & \cellcolor[HTML]{BCEBBC}Yes \\
\textbf{Regulator} & \cellcolor[HTML]{FFAFAF}No  & \cellcolor[HTML]{FFAFAF}No  & \cellcolor[HTML]{BCEBBC}Yes \\ \hline
\end{tabular}
}
\end{table}

\begin{figure}[h!]
\centering
\includegraphics[scale=0.52]{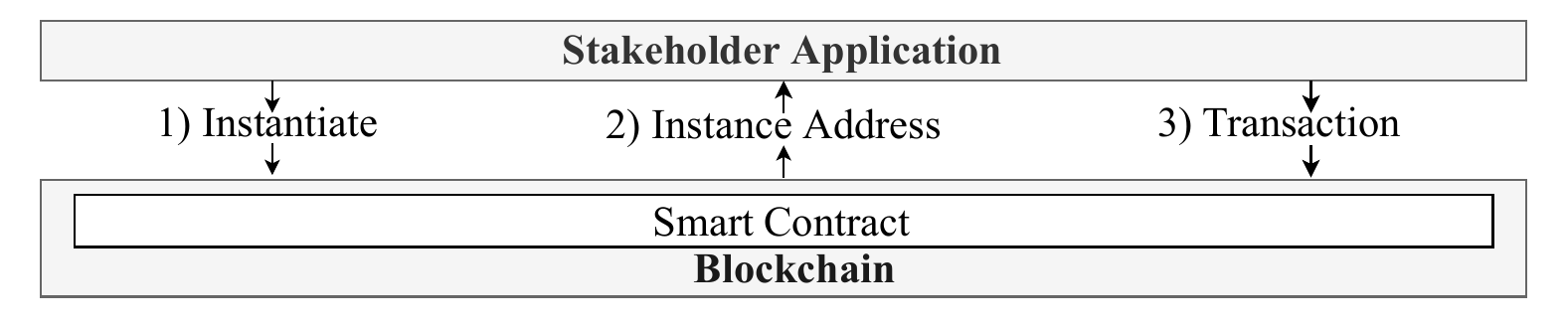}
\caption{Smart contract instance and transaction validation}
\label{fig:instance-smart-contract.pdf}
\end{figure}

Figure \ref{fig:instance-smart-contract.pdf} shows the steps for creating an instance and sending transactions. After uploading the contract to the network, an instance of the contract is created among the stakeholders with the definition of who sends and receives a transaction. The stakeholder (sender) sends a transaction to the network informing the instance address and the transaction data described in Section \ref{sec:overview-smart-contracts}. For example, the doctor sends a transaction to the network informing the instance address and the prescription data (i.e., patient personal information, medication, and diagnosis). The contract method will verify the transaction's validity by checking if the sender is the same as defined by the instance. To be added to the blockchain, transactions will follow the consensus steps of the underlying blockchain. If it is a valid transaction, the current contract state will be updated with the transaction data. 

\begin{figure*}[h!]
\centering
\includegraphics[scale=0.44]{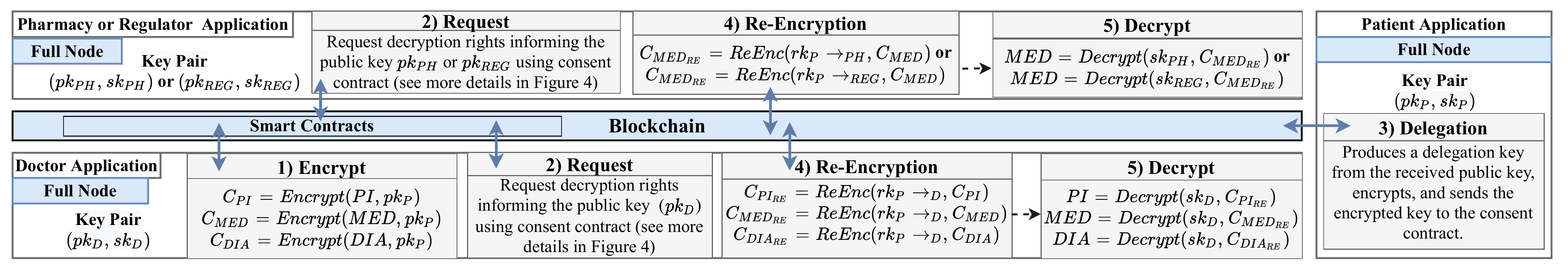}
\caption{Our decentralized data governance framework with support for PRE mechanism, consent management, data provenance, and privacy.}
\label{fig:architecture.pdf}
\end{figure*}




\subsection{Proxy Re-Encryption Mechanism}

In this work, we use the proxy re-encryption technique to ensure data privacy. In this way, stakeholders can decrypt prescription data only with the patient's consent via the delegation mechanism. 
Table \ref{tab:privacy-levels} shows the information that stakeholders can consult after the patient's permission, and the diagram in Figure \ref{fig:architecture.pdf} shows the architecture with the PRE operations:



\begin{enumerate}



\item From appointment with the patient, the doctor creates a prescription containing the items: personal information (PI), medication (MED), and diagnosis (DIA) for future analysis. Before sending the prescription to the $create\_prescription$ smart contract method and being stored in the contract state, the prescription items are encrypted by the doctor application separately using the patient's public key ($pk_P$). \emph{Therefore, the patient has flexibility and can consent to data sharing}.



\item For the doctor, pharmacy, or regulator to analyze any item in the prescription, it will be necessary to request decryption rights from the patient informing their respective public key ($pk$).

\item If the patient agrees with the request, the patient's application will generate a delegation key following the privacy filters in Table \ref{tab:privacy-levels}. The delegation key will be encrypted and sent to the \textit{set\_consent} contract method.

\item In the stakeholder application, the proxy will perform the re-encryption ($RE$) operation using the respective delegation key and the allowed prescription item, as shown in Table \ref{tab:privacy-levels}. The doctor has access to all the prescription data. For this, the proxy will perform the re-encryption for personal information $C_{PI}$, medication $C_{MED}$ and diagnosis $C_{DIA}$. The pharmacy and the regulator can only re-encrypt the prescribed medication.





\item After the re-encryption step, stakeholders can decrypt the item with their respective private key ($s_k$) and analyze the information allowed by the patient.



\end{enumerate}
Note that sensitive prescription items are encrypted before being stored in the blockchain. In this way, records are private and immutable. Other organizations will only be able to decrypt the information with the patient's permission. The proxy re-encryption mechanism is a privacy software module implemented in the stakeholder application to act on confidential data sharing operations.

\textbf{Note about proxy}: proxy is a software that only re-encrypts information. The proxies do not store any private keys and do not see any message from the ciphertexts. From their perspective, they only see an incoming ciphertext and the result after re-encryption, which is also a ciphertext.

\subsubsection{Consent mechanism and delegation key}
\label{sec:request-smart-contract-delegation-key}

\begin{figure}[h!]
\centering
\includegraphics[scale=0.45]{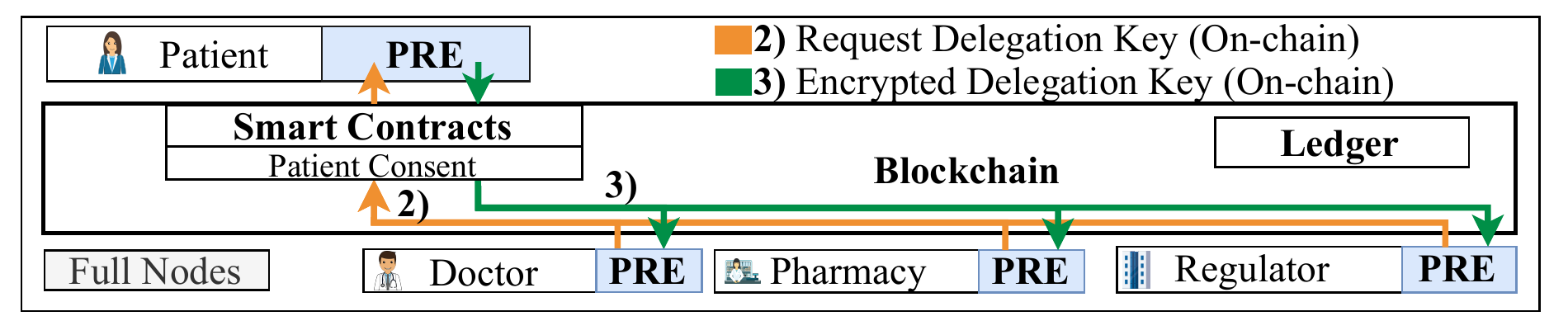}
\caption{On-chain request mechanism (step 2) and sending the encrypted delegation key (step 3) using smart contracts }
\label{fig:request-delegation.pdf}
\end{figure}

Figure \ref{fig:request-delegation.pdf} represents steps 2 and 3 of Figure \ref{fig:architecture.pdf} where each stakeholder is a full node (i.e., containing the PRE operations and blockchain). In step 2, the stakeholder sends a request to the patient through a consent contract. In step 3, the patient will create and encrypt the delegation key using the stakeholder's public key. In this way, requests are transparent to all network participants, and only the stakeholder can decrypt the delegation key.

\subsection{How does the proposed solution meet the objectives?}

\textbf{Objective 1} focuses on providing transparency to the prescription system while protecting patient's privacy — our solution stores the patients' data on the blockchain but in an encrypted form. The prescription data is made available to other parties after the patient's consent. 

\textbf{Objective 2} focuses on governing data usage - our solution tracks the data access requests of consumers and permissions of data owners through a smart contract and immutable ledger. Therefore, data owners can have visibility into their data and its usage. We understand that a malicious data consumer may access the patient's data with their permission and then post it on a black market or other digital platforms. We plan to investigate digital watermarking and steganography in our future work to overcome this problem~\cite{zhao2018bmcprotector}.

\textbf{Objective 3} aims to prevent illegal sales of drugs — our solution includes a regulator in the prescription system, thereby providing transparency and accountability. By allowing the regulator to track the flow of goods, our solution reduces illegal drug sales. We acknowledge that the pharmacies may acquire drugs from outside the system (without leaving a digital record). In such cases, patients can report pharmacies that sell drugs unlawfully to the regulators in return for an incentive, following the CrimeStoppers model (see Section~\ref{sec:disc}).

\section{Threat Analysis}
\begin{figure*}[h!]
\centering
\includegraphics[scale=0.45]{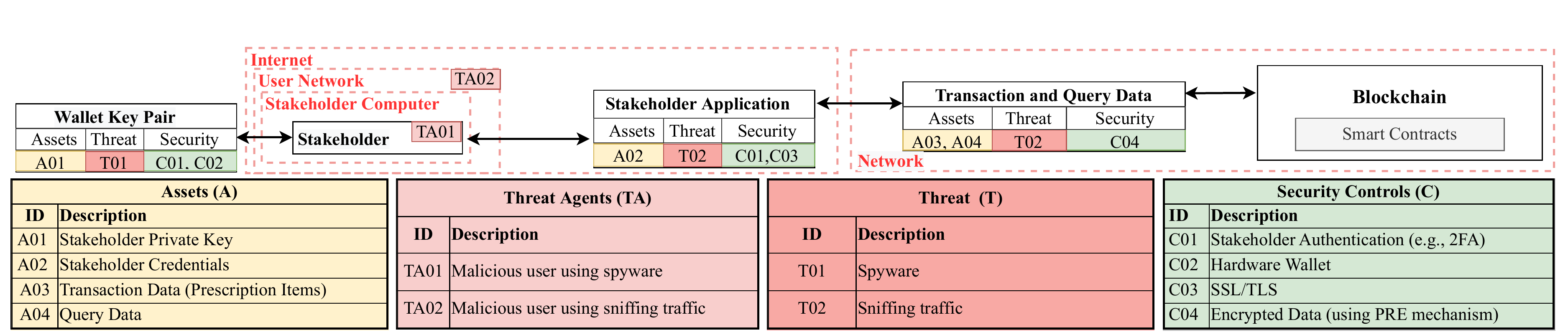}
\caption{Threat analysis for decentralized electronic prescription system}
\label{fig:threat-model.pdf}
\end{figure*}

To reduce the model's security risks, we list potential threats that affect electronic prescription system data integrity, availability, and privacy. The goal is to identify and mitigate external and internal threats through countermeasures. The diagram in Figure \ref{fig:threat-model.pdf} shows the Assets (A), Threats agents (TA), Threats (T), and Security Controls (C).

\subsection{Assets}
Assets represent values in the system. In the prescription model, private keys, stakeholder credentials, transaction data are assets.

\begin{itemize}
\item Stakeholder Private Key (A01): used to create and sign transactions;
\item Stakeholder Credentials (A02): user authentication in the application;
\item Transaction Data (A03): information about the patient personal information, medications, and diagnosis;
\item Query Data (A04): query the contract state data;
\end{itemize}

\subsection{Threat Agents}
Threat agents are malicious users who aim to obtain assets to harm the security and privacy of the system. These users can be within the organization, impersonating a doctor, pharmacist, patient or regulator. These threat agents seek to intercept packets on the network (using packet analyzers) and disrupt the functionality of applications and services.

\begin{itemize}
\item Threat Agent (Insider): uses the stakeholder's computer to try to capture the private key;
\item Threat Agent (External): uses tools to analyze packets (e.g., sniffing traffic) in the stakeholder and consensus networks. The primary purpose is to capture information about users and transactions;
\end{itemize}

\subsection{Threats}
Threats are ways for an agent to get an asset and negatively affect the system (e.g., get private keys, credentials and sensitive prescription information).

\begin{itemize}
\item Spyware: capture the stakeholder's private key to create fraudulent transactions (e.g., prescriptions, amount of medication sold or supplied);

\item Sniffing traffic: obtain credentials to access the stakeholder application and examine valuable information (e.g., patient's personal information, medications, and diagnosis);

\end{itemize}

\subsection{Security Controls (Countermeasures)}
To reduce the risk of compromising assets, we have proposed minimal countermeasures to ensure security, privacy, and system availability:

\begin{itemize}
    \item Stakeholder authentication and hardware wallet (C01 and C02): prevent threat agents from obtaining stakeholder private keys. The private key must be kept in an offline wallet or on a hardware wallet (HW Wallet) and accessed using an authentication system, for example, two-factor authentication (2FA);
    
    \item Stakeholder authentication and SSL/TLS encryption (C01 and C03): access to the stakeholder application must be through an authentication mechanism, and the communication between user and application must be encrypted (HTTPS) with SSL/TLS certificate;
    
    \item Encrypted data using PRE mechanism (C04): data sent and consulted on the network in encrypted form;
    

\end{itemize}

\section{Proof-of-Concept Implementation and Evaluation}

\subsection{Privacy: Proxy Re-Encryption}

\textbf{Evaluation Goals:} To understand the overhead and feasibility of PRE operations, we evaluate memory allocation and execution time tests for steps in Figure \ref{fig:architecture.pdf}. We evaluated encrypting (step 1), creating a delegation key (step 3), re-encryption (step 4), and decrypting (step 5). 

\begin{figure}[h!]
\centering
\includegraphics[scale=0.46]{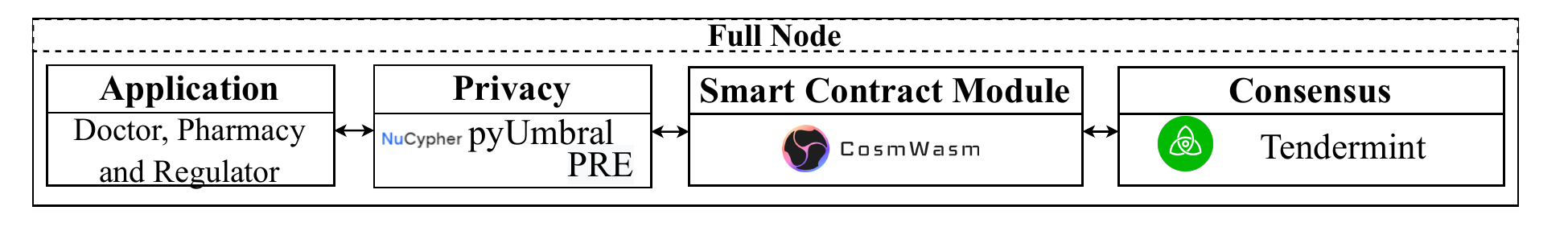}
\caption{Implementation using NuCypher pyUmbral PRE for privacy and CosmWasm smart contract module}
\label{fig:tecnologies-used.pdf}
\end{figure}

\textbf{Evaluation Setup and Methodology:} The PRE evaluation programs and scripts were implemented in Python programming language using NuCypher pyUmbral PRE technology, an open-source implementation that uses the \textit{secp256k1} elliptic curve \cite{github-pyumbral-PRE}. We use the module \textit{tracemalloc}, a tool to trace memory blocks allocated by the evaluation program during the execution of the operations. For execution time, we use the \textit{time} module to calculate the difference between the start and end of each operation. All software created for evaluation is available on GitHub \cite{github-repo}. Figure \ref{fig:tecnologies-used.pdf} shows the NuCypher pyUmbral PRE technology used as a privacy mechanism and CosmWasm a smart contract module in the Cosmos ecosystem. For comparison we implement the contracts using the Ethereum platform and will be discussed in the section \ref{sec:ethereum-implementation}.

To identify realistic file sizes for medications and dosage prescriptions, we used the English Prescribing Dataset \cite{EnglishPrescribingDataset(EPD)}. Prescription items used for evaluation are represented in separate text files with different sizes ranging from 0.43 Kilobyte (kB) to 0.82 kB for personal information, 0.24 kB to 0.53 kB for medication and 2.18 kB to 8975.74 kB $\approx$ 8.76 Megabyte (MB) for diagnosis. While the file sizes are inferred from \cite{EnglishPrescribingDataset(EPD)}, file contents are randomly generated by the evaluation software. In total, 1000 iterations were performed for different file sizes. We used a Linux virtual machine with an Intel Core i7-10510U 1.80GHz (Dual-Core) processor and 6GB of RAM for the evaluation.

Table \ref{tab:prescription-data} presents information about the maximum (max.), minimum (min.), and average (avg.) size of each prescription item. 


\begin{table}[]
\centering
\caption{Maximum, minimum and average files size with prescription items}
\label{tab:prescription-data}
\begin{tabular}{cccc}
\hline
\multicolumn{4}{c}{\textbf{Prescription Data (kB)}}        \\ \hline
\textbf{File Size} &
  \textbf{\begin{tabular}[c]{@{}c@{}}Personal \\ Info\end{tabular}} &
  \textbf{\begin{tabular}[c]{@{}c@{}}Medication \\ and Dosage\end{tabular}} &
  \textbf{Diagnosis} \\
\textbf{Min.} & 0.43 & 0.24 & 2.18                         \\
\textbf{Max.} & 0.82 & 0.53 & 8975.74 ($\approx$ 8.76 MB)  \\
\textbf{Avg.} & 0.62 & 0.39 & 4538.57  ($\approx$ 4.43 MB) \\ \hline
\end{tabular}
\end{table}

\begin{figure}[h!]
\centering
\includegraphics[scale=0.30]{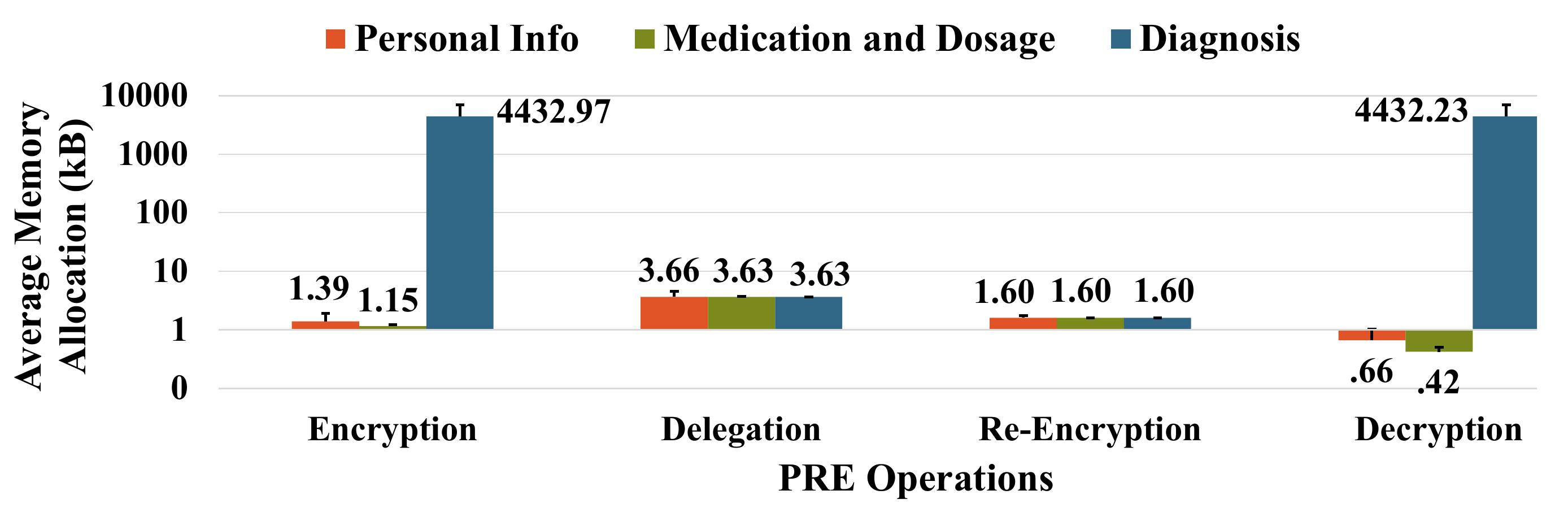}
\caption{Average memory allocation by \textit{tracemalloc} in PRE operations}
\label{fig:memory-allocation.pdf}
\end{figure}

\begin{figure}[h!]
\centering
\includegraphics[scale=0.30]{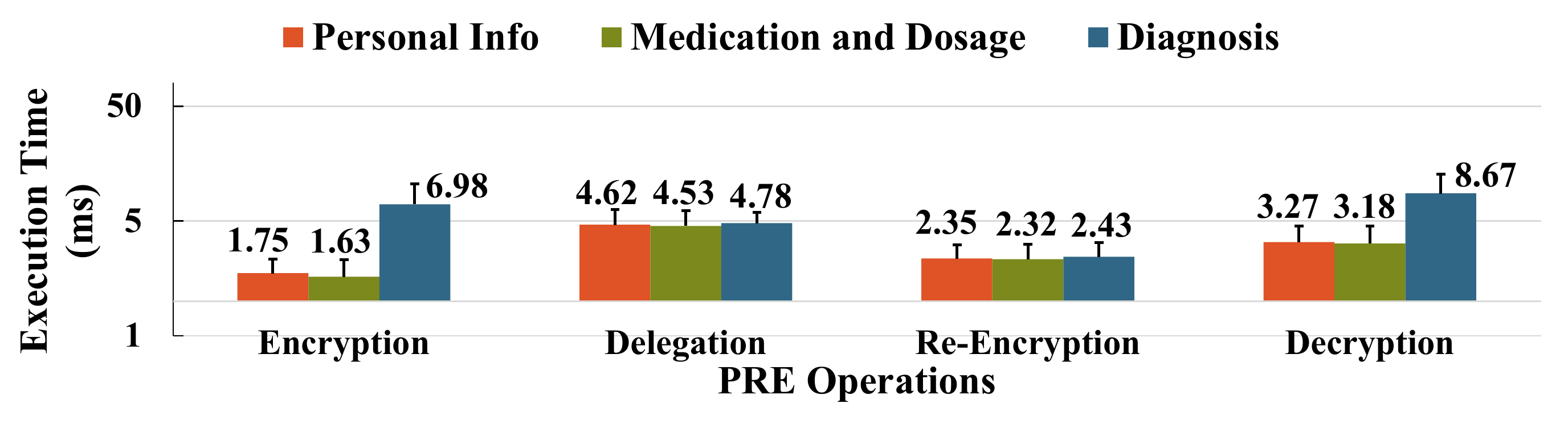}
\caption{Average execution time in PRE operations}
\label{fig:execution-time.pdf}
\end{figure}

\subsubsection{\textbf{Memory Allocation Evaluation}}

Figure \ref{fig:memory-allocation.pdf} shows the average memory allocated for application-level PRE operations using the data files in Table \ref{tab:prescription-data}. In all operations, most memory blocks allocated are for asymmetric cryptography operations using the \textit{secp256k1} curve. In particular, pyUmbral uses OpenSSL via \textit{Cryptography.io} library. We highlight the key findings below:

\begin{itemize}
    \item The diagnosis data's encryption consumes significant memory since it contains a large amount of data compared to other data items (see Table \ref{tab:prescription-data}). The diagnosis data requires an average allocation of 4432.97 kB ($\approx$ 4.32 MB) with std. of 2524.97 kB ($\approx$ 2.46 MB), while for personal information and medication, it was 1.39 kB and 1.15 kB with std. of 0.55 kB and 0.08 kB, respectively, as shown in Figure~\ref{fig:memory-allocation.pdf}.
    \item In the delegation stage (step 3 in Figure \ref{fig:architecture.pdf}), which is performed on the patient's application, the average memory allocation was around 3.66 kB (with std. of 0.91 kB) for personal information,  medication, and diagnosis. Similarly, in the re-encryption stage (step 4), the average allocated memory was around 1.60 kB (with std. of 0.01 kB). 
    \item Similar to the encryption stage, the decryption (step 5) also consumes significant memory due to the execution of computationally-intensive cryptography operations.
\end{itemize}





\subsubsection{\textbf{Execution Time Evaluation}}

Figure \ref{fig:execution-time.pdf} shows the average execution time for application-level PRE operations using the data files in Table \ref{tab:prescription-data}.

To encrypt the diagnostic data (step 1), it took an average of 6.98 milliseconds (ms) with a standard deviation of 3.56 ms for files ranging from 2.18 kB to 8975.74 kB $\approx$ 8.76 Megabyte (MB). The medication and dosage data took an average of 1.63 ms with a standard deviation of 0.67 ms for data files ranging from 0.24 kB to 0.53 kB. For personal information, it took 1.75 ms with a standard deviation of 0.59 ms for data files from 0.43 kB to 0.82 kB.

Similar to the average memory allocation evaluation, there were slight variations in the average processing times in the delegation and re-encryption stage for the prescription items. The delegation stage (step 3) took an average of 4.78 ms (with std. of 1.17 ms) for diagnosis data, 4.53 ms (with std. of 1.65 ms) for medication and dosage, and 4.62 ms (with std. of 1.66 ms) for personal information.  In the re-encryption operation (step 4), the average execution time was 2.43 ms (with std. of 0.81) for diagnosis, 2.32 ms (with std. of 0.82 ms) for medication and dosage, and 2.35 ms (with std. of 0.76) for the patient's personal information. In the decrypt stage (step 5), the item that obtained the highest execution average was the diagnosis with 8.67 ms (with std. of 4.05 ms). In comparison, medication and dosage took 3.18 ms, and the patient's personal information took 3.27 ms with std. of 1.36 ms and 1.25 ms, respectively.

\emph{These results show PRE operations' cost of memory allocation and execution time is relative to the data size. In our evaluation, even with text files with sizes in Megabyte (see Table \ref{tab:prescription-data}), the operations did not exceed 50 ms to be executed. In this sense, our proposed framework protects privacy and manages consent with a low operational overhead.  We also believe PRE operations can run on platforms like Raspberry Pi or mobile phones.}



\subsection{Smart Contract: CosmWasm Implementation}

A test network called \textit{Uni Junø} network  \cite{github-testnet-uni-juno} with 30 validator nodes was used to evaluate the transaction time of the encrypted prescription items (after encryption step in Figure \ref{fig:architecture.pdf}). We measure the time required to validate a transaction by the contract’s \textit{create\_prescription} method. The steps to automate the sending of transactions to the network were implemented in a shell script. All software and contracts developed for model evaluation are available on GitHub \cite{github-repo}.


\textbf{Transaction time for Smart Contracts in CosmWasm:} Table \ref{tab:evaluation-cosmwasm} shows the transaction validation time for each prescription ranging from 0.92 kB to 130.50 kB containing all items (i.e., patient's personal information, medication, and diagnosis). Time refers to the Tendermint consensus process with inclusion in a block. For \textit{Uni Junø} network, the average block generation time is 6.13 seconds (informed by \textit{Uni Junø} block explorer \cite{github-testnet-uni-juno}). Table \ref{tab:evaluation-cosmwasm} shows the maximum (max.), minimum (min.), average (avg.) and standard deviation (std) time for 300 iterations. On average, the time for a transaction to be validated by contract method took 2.69 seconds, with the maximum and minimum time being 6.26 seconds and 1.50 seconds, respectively. The variation in transaction time is due to the consensus delay, including peer-to-peer messaging between validator nodes.

\begin{table}[]
\centering
\caption{Transaction validation time for the create\_prescription contract method using CosmWasm implementation with \textit{Uni Junø} testnet}
\label{tab:evaluation-cosmwasm}
\begin{tabular}{ccclc}
\hline
                                                                           & \multicolumn{4}{c}{\textbf{Transaction time (in seconds)}}              \\ \hline
\textbf{\begin{tabular}[c]{@{}c@{}}Number of \\ Transactions\end{tabular}} & \textbf{Max.} & \textbf{Min.} & \textbf{Avg.}            & \textbf{Std.} \\
300                                                                        & 6.26          & 1.50          & \multicolumn{1}{c}{2.69} & 0.71         \\ \hline
\end{tabular}
\end{table}



\textbf{A note on CosmWasm's transaction cost:} At its current stage, the CosmWasm platform is an smart contract module that can plug into the Cosmos SDK. The minimum price of each transaction is determined by each chain. We cannot provide the transaction price converted to USD because, currently, CosmWasm does not have any tradeable token. For the experiment, we used \textit{0.025ujunox} symbolic tokens as a test network parameter for sending transactions (tokens were obtained through a network faucet server \cite{github-testnet-uni-juno}).

\begin{table}[]
\centering
\caption{Contract deploy information using the Ethereum implementation and the Ropsten testnet}
\label{tab:ethereum-contract-deploy}
\begin{tabular}{@{}lccccc@{}}
\toprule
\multirow{2}{*}{\textbf{Smart Contract}} & \multicolumn{1}{c|}{\multirow{2}{*}{\textbf{\begin{tabular}[c]{@{}c@{}}Data Size \\ (bytes)\end{tabular}}}} & \multicolumn{3}{c|}{\textbf{Block Mining Time (s)}}                                                          & \multirow{2}{*}{\textbf{\begin{tabular}[c]{@{}c@{}}Txn. Fee\\  (ETH)\end{tabular}}} \\ \cmidrule(lr){3-5}
                                         & \multicolumn{1}{c|}{}                                                                                       & \multicolumn{1}{c|}{\textbf{Min.}} & \multicolumn{1}{c|}{\textbf{Max.}} & \multicolumn{1}{c|}{\textbf{Avg.}} &                                                                                     \\ \midrule
Prescription                             & 3021                                                                                                        & 1                                  & 105                                & 11.69                              & 0.00177497                                                                          \\
Report                                   & 1419                                                                                                        & 1                                  & 60                                 & 10.44                              & 0.00088486                                                                          \\
Sales                                    & 2245                                                                                                        & 1                                  & 50                                 & 11.97                              & 0.00136244                                                                          \\
Medication Control                       & 1771                                                                                                        & 1                                  & 50                                 & 11.54                              & 0.00110506                                                                          \\
Consent                                  & 3552                                                                                                        & 1                                  & 87                                 & 13.27                              & 0.00205878                                                                          \\
Reward                                   & 532                                                                                                         & 1                                  & 58                                 & 13.18                              & 0.00042597                                                                          \\ \bottomrule
\end{tabular}
\end{table}


\subsection{Smart Contract: Ethereum Implementation}
\label{sec:ethereum-implementation}
\begin{table*}[]
\centering
\caption{prescription and consent contract methods with fee per transaction in ethereum implementation with ropsten testnet}
\label{tab:smart-contract-methods-ethereum-1}
\begin{tabular}{@{}ccccccccc@{}}
\toprule
\multirow{2}{*}{\textbf{\begin{tabular}[c]{@{}c@{}}Steps in \\ Figure \ref{fig:architecture.pdf}\end{tabular}}} &
  \multirow{2}{*}{\textbf{\begin{tabular}[c]{@{}c@{}}Event/Contract \\ Method\end{tabular}}} &
  \multirow{2}{*}{\textbf{\begin{tabular}[c]{@{}c@{}}Hexadecimal \\ Data Size (bytes)\end{tabular}}} &
  \multirow{2}{*}{\textbf{Invoked by}} &
  \multicolumn{1}{c|}{\multirow{2}{*}{\textbf{Smart Contract}}} &
  \multicolumn{3}{c|}{\textbf{Block Mining Time (s)}} &
  \multirow{2}{*}{\textbf{Txn. Fee (ETH)}} \\ \cmidrule(lr){6-8}
 &
   &
   &
   &
  \multicolumn{1}{c|}{} &
  \multicolumn{1}{c|}{\textbf{Min.}} &
  \multicolumn{1}{c|}{\textbf{Max.}} &
  \multicolumn{1}{c|}{\textbf{Avg.}} &
   \\ \midrule
1 &
  create\_prescription &
  7012 &
  Doctor &
  Prescription &
  1 &
  58 &
  12.17 &
  0.00159011 \\
2 &
  request\_delegation &
  100 &
  Doctor and Pharmacy &
  Consent &
  1 &
  57 &
  12.69 &
  0.00012677 \\
3 &
  set\_consent &
  612 &
  Patient &
  Consent &
  1 &
  48 &
  10.43 &
  0.00022021 \\ \bottomrule
\end{tabular}
\end{table*}


\begin{table*}[]
\centering
\caption{sales, report, reward and medication control contract methods with fee per transaction in ethereum implementation with ropsten testnet}
\label{tab:smart-contract-methods-ethereum-2}
\begin{tabular}{@{}lccccccc@{}}
\toprule
\multirow{2}{*}{\textbf{\begin{tabular}[c]{@{}l@{}}Event/Contract \\ Method\end{tabular}}} &
  \multirow{2}{*}{\textbf{\begin{tabular}[c]{@{}c@{}}Hexadecimal \\ Data Size (bytes)\end{tabular}}} &
  \multirow{2}{*}{\textbf{Invoked by}} &
  \multicolumn{1}{c|}{\multirow{2}{*}{\textbf{Smart Contract}}} &
  \multicolumn{3}{c|}{\textbf{Block Mining Time (s)}} &
  \multirow{2}{*}{\textbf{Txn. Fee (ETH)}} \\ \cmidrule(lr){5-7}
 &
   &
   &
  \multicolumn{1}{c|}{} &
  \multicolumn{1}{c|}{\textbf{Min.}} &
  \multicolumn{1}{c|}{\textbf{Max.}} &
  \multicolumn{1}{c|}{\textbf{Avg.}} &
   \\ \midrule
sell\_medication &
  356 &
  Pharmacy &
  Sales &
  1 &
  85 &
  10.83 &
  0.00013032 \\
create\_report &
  676 &
  Patient &
  Report &
  1 &
  55 &
  11.78 &
  0.00021107 \\
send\_reward &
  36 &
  Regulator &
  Reward &
  1 &
  52 &
  11.24 &
  0.00006627 \\
supply\_medications &
  36 &
  Regulator &
  Medication Control &
  1 &
  61 &
  13.86 &
  0.00009519 \\
update\_medications\_sold &
  36 &
  Pharmacy &
  Medication Control &
  1 &
  102 &
  14.18 &
  0.00009589 \\ \bottomrule
\end{tabular}
\end{table*}
We measure the transaction (txn) fee and the average block mining time in the Ethereum platform using the Ropsten test network. We have implemented the smart contracts using Solidity and used the Remix platform to deploy and interact with contract methods. MetaMask wallet was used to obtain transaction details (e.g., transaction fee and data size), and Etherscan to monitor transactions and blockchain information.

Table \ref{tab:ethereum-contract-deploy} presents the contracts with the size in bytes with the costs necessary for the deployment on the network (from MetaMask wallet). We performed 100 iterations for each contract and method to calculate the block's average mining time. The average mining time remained between 10.43 seconds and 14.18 seconds. As a comparison, \textit {Junø Scan} a block explorer used to evaluate CosmWasm transactions calculates the average time of 6.13 seconds for block generation (commit).

Table \ref{tab:smart-contract-methods-ethereum-1} shows the information about sending transactions for the contract methods: the doctor creates a prescription, requests delegation, and for the patient to authorize access to the information according to the steps shown in Figure \ref{fig:architecture.pdf}.
Table \ref{tab:smart-contract-methods-ethereum-2} shows transaction information about medication sales, creating a report, sending rewards, supplying medications, and updating medication sold.

\textbf{Note on Blockchain-agnosticism:} The evaluation with CosmWasm and Ethereum shows that our data governance framework is agnostic to the underlying blockchain platform. One can implement our data governance framework on any blockchain platform that supports smart contracts.

\section{Discussion: CrimeStopper model for E-prescription}
\label{sec:disc}
The proposed decentralized e-prescription system can provide transparency while preserving patients' privacy through a secure and scalable consent management solution. Besides, the inclusion of the regulatory board into the architecture prevents pharmacies from selling drugs illegally. Here, illegal drug sales happen because patients may not be willing to consult the doctor. Instead, they visit pharmacies directly to get medications, including antibiotics. Therefore, the architecture proposed in this paper cannot fully mitigate the illegal sales since the patient may collude with a pharmacy to get the drugs they want. In this section, we discuss a possible solution using a CrimeStopper model.

The CrimeStopper model relies on the community members to solve crimes. Law enforcement agencies provide a reward when community members report crimes or provide essential clues to solve a crime. We propose integrating the CrimeStopper model in the architecture presented in Figure 3 to encourage the community members to report pharmacies that sell illegal drugs. Such a solution would expose dishonest pharmacies and reduce unlawful sales significantly as the pharmacies may prefer to behave honestly to protect their business. For such a solution to work, 1) A significant reward for the patient that reports illegal sales with evidence, 2) A significant penalty, including the cancellation of license for repeated offenders, for the pharmacies that get caught selling illegal drugs. We plan to evaluate the CrimeStopper model for e-prescription using game theory in our future work to weigh its effectiveness.

\section{Conclusions}
Real-world multi-stakeholder applications such as e-prescription and supply chain deal with digital and sensitive data, demanding privacy protection, consent management, data provenance, and transparency. We have presented a decentralized data governance framework for e-prescription that uses proxy re-encryption and smart contracts to let data owners control and manage their data through a trusted and transparent blockchain platform. We have shown how the data owners can record all the access requests and consents in an immutable ledger to monitor data lineage. Our proof-of-concept implementation uses CosmWasm, Ethereum, and pyUmbral proxy re-encryption library to assess the feasibility and performance. Our evaluation results show that the proposed architecture can protect data owners' privacy and govern sensitive data access with minimal overhead. We believe that our data governance framework is beneficial to all multi-stakeholder applications that deal with sensitive and private digital data.

\bibliographystyle{unsrtnat}
\bibliography{references}

\end{document}